\documentclass[11pt]{article}
\usepackage{amsmath}
\usepackage{amssymb}
\usepackage{amsmath,amsthm,epsfig,euscript,array,cite,cancel}
\usepackage{mathtools}
\usepackage[linktocpage]{hyperref}
\usepackage{ytableau}

\newcommand{\be}{\begin{equation}}
\newcommand{\bea}{\begin{eqnarray}}
\newcommand{\ee}{\end{equation}}
\newcommand{\eea}{\end{eqnarray}}

\begin{document}

\newcommand{\ds}{\displaystyle}

\topmargin -1cm \oddsidemargin=0.25cm\evensidemargin=0.25cm
\setcounter{page}0
\renewcommand{\thefootnote}{\fnsymbol{footnote}}

\begin{titlepage}
\date{}

\title{\bf Fermionic higher-spin triplets in AdS }
\maketitle

\begin{center}

\author{Alessandro Agugliaro $^{a}$\footnote{e-mail: {\tt alessandro.agugliaro@fi.infn.it}}},
\author{Francesco Azzurli $^b$\footnote{e-mail: {\tt  francesco.azzurli@mib.infn.it}  }} and
\author{Dmitri Sorokin$^c$\footnote{e-mail: {\tt  dmitri.sorokin@pd.infn.it }} }
 \vskip .4in {$^a$ \it Dipartimento di Fisica, Universit\'a di Firenze and INFN Sezione di Firenze, Via G. Sansone 1,
50019 Sesto Fiorentino, Italy } \\
\vskip .2in {$^b$ \it Dipartimento di Fisica, Universit\`a degli studi di Milano-Bicocca and INFN, \\ Sezione di Milano--Bicocca, Piazza della Scienza 3, 20126 Milano, Italy} \\
\vskip .2in {$^c$ \it INFN, Sezione di Padova, via F. Marzolo 8, 35131 Padova, Italia}
\vskip .8in
\begin{abstract}
\noindent
We derive a metric-like Lagrangian and equations of motion, in $AdS$ space, for multiplets of fermionic fields with spin ranged from $\frac 12$ to $s$, from their frame-like formulation.
\end{abstract}

\end{center}


\thispagestyle{empty}
\end{titlepage}

\renewcommand{\thefootnote}{\arabic{footnote}}
\setcounter{footnote}0

\section{Introduction}
Systems of massless higher spin fields which are transformed under a reducible representation of the Lorentz group have attracted a great deal of attention (see e.g. \cite{Ouvry:1986dv,Bengtsson:1986ys,Bellon:1986ki,Henneaux:1988,Pashnev:1989gm,Francia:2002pt,Sagnotti:2003qa,Barnich:2005ga,Buchbinder:2006eq,Buchbinder:2007ak,Fotopoulos:2007yq,Fotopoulos:2008ka,Sorokin:2008tf,Fotopoulos:2009iw,Fotopoulos:2010nj,Francia:2010qp,Francia:2012rg,Bekaert:2010hk,Skvortsov:2010nh,Campoleoni:2012th,Asano:2012qn,Asano:2013rka,Bekaert:2015fwa}), since, as was shown 30 years ago \cite{Ouvry:1986dv,Bengtsson:1986ys}, they arise in a tensionless limit of string theory in a flat background and may, therefore, shed light on a possible relation between string theory and higher spin gauge theory. The simplest of these systems consists of three symmetric tensor fields of rank $s$, $s-1$ and $s-2$. This motivated Francia and Sagnotti \cite{Francia:2002pt} to call them \emph{higher-spin triplets}. The simplest massless bosonic triplets form reducible multiplets of physical fields with even or odd spins (or helicities) running, respectively, from 0 or 1 to $s$, while the fermionic triplets consist of physical fields with half-integer spins running from 1/2 to $s$ \footnote{The conventional notion of spin and helicity are associated with irreducible representations of the four-dimensional Poincar\'e group. In what follows we  will formally use this terminology also in higher-dimensional theories associating the ``spin" with the rank of symmetric tensor-spinors which transform under irreducible representations of the $SO(1,D-1)$ Lorentz group.}. More complicated systems involve tensor fields of mixed symmetry. Since 1986 higher-spin triplets have been studied from different perspectives (see e.g. \cite{Ouvry:1986dv,Bengtsson:1986ys,Bellon:1986ki,Henneaux:1988,Pashnev:1989gm,Francia:2002pt,Sagnotti:2003qa,Barnich:2005ga,Metsaev:2005ar,Buchbinder:2006eq,Buchbinder:2007ak,Fotopoulos:2007yq,Metsaev:2007rn,Fotopoulos:2008ka,Sorokin:2008tf,Buchbinder:2008ss,Fotopoulos:2009iw,Fotopoulos:2010nj,Francia:2010qp,Francia:2012rg,Bekaert:2010hk,Skvortsov:2010nh,Campoleoni:2012th,Metsaev:2012uy,Asano:2012qn,Asano:2013rka,Bekaert:2015fwa}).  For a review and references on difference aspects of higher spin theory see e.g. \cite{Vasiliev:2004qz,Sorokin:2004ie,Bekaert:2005vh,Sagnotti:2005ns,Francia:2006hp,Fotopoulos:2008ka,Bekaert:2010hw,Gaberdiel:2012uj,Giombi:2012ms,Sagnotti:2013bha,Didenko:2014dwa,Rahman:2015pzl}).

Fermionic triplets of tensor-spinor fields were introduced in \cite{Bellon:1986ki} and independently in \cite{Francia:2002pt}. Their origin from a tensionless limit of a Ramond-Neveu-Schwarz string was studied in detail in \cite{Sagnotti:2003qa}.  The Lagrangian formulations (i.e. the actions and equations of motion) for bosonic triplets are known in flat and anti de Sitter spaces, both in metric-like and frame-like formulations of the higher spin fields. For the fermionic triplets, however, the metric-like actions and equations of motion have been constructed only in flat space-time, while the generalization of this construction to the anti-de Sitter spaces encountered obstacles \cite{Sagnotti:2003qa,Buchbinder:2007ak} and have not been fulfilled by now. On the other hand, in the frame-like formalism the Lagrangian description of reducible fermionic higher-spin systems in AdS was constructed in \cite{Sorokin:2008tf} and it was outlined therein that the metric-like Lagrangian formulation of the fermionic triplets in AdS can be obtained from the frame-like one by a certain redefinition of fields.

The purpose of this note is to accomplish this goal and to derive an explicit form of the metric-like action, equations of motion and local symmetries of the fermionic triplets in AdS from their frame-like counterparts. In passing we will clarify a subtlety regarding a physical spin-1/2 field one should deal with when relating the reducible fermionic frame-like higher-spin system to the metric-like fermionic triplet. A motivation for presenting these results is that having got the gauge invariant metric-like Lagrangian formulation of the fermionic (and bosonic) higher-spin triplets in AdS one can perform its BRST analysis (already carried out in the bosonic case in \cite{Sagnotti:2003qa}) with the aim of understanding whether and how these systems may be obtained by taking a tensionless limit of a String Theory in an AdS background.

Our main notation and conventions are given in the Appendix.

\section{Fermionic triplets in flat space-time}

\subsection{Metric-like formulation \label{sec:MLFTriplets}}

In the metric-like formulation in D-dimensional space-time a fermionic higher-spin triplet consists of three symmetric tensor-spinor fields
$\Psi_\alpha^{\mu_{1}...\mu_{r}}$, $\chi^{\mu_{1}...\mu_{r-1}}_\alpha$ and $\lambda_\alpha^{\mu_{1}...\mu_{r-2}}$, where $\mu_i=0,1,\ldots, D-1$ are $D$-dimensional space-time indices, $\alpha$ is a spinor index which we will usually skip, and $r=s-\frac{1}{2}$ with $s$ denoting the highest spin in the spectrum of the triplet fields. In flat space-time they satisfy the following equations of motion \cite{Francia:2002pt}
\begin{align}
\cancel{\partial}\Psi^{\mu_{1}...\mu_{r}}+i\partial^{(\mu_{1}}\chi^{\mu_{2}...\mu_{r})} & =0,\nonumber \\
\partial_\nu\Psi^{\nu\mu_{1}...\mu_{r-1}}-\partial^{(\mu_{1}}\lambda^{\mu_{2}...\mu_{r-1})}+i\cancel{\partial}\chi^{\mu_{1}...\mu_{r-1}} & =0,\label{ML-FTripletEqs}\\
\cancel{\partial}\lambda^{\mu_{1}...\mu_{r-2}}+i\partial_\nu\chi^{\nu\mu_{1}...\mu_{r-2}} & =0\,,\nonumber
\end{align}
which are invariant under the gauge transformations
\begin{align}
\delta\Psi^{\mu_{1}...\mu_{r}} & =\partial^{(\mu_{1}}\Lambda^{\mu_{2}...\mu_{r})},\nonumber \\
\delta\chi^{\mu_{1}...\mu_{r-1}} & =i\cancel{\partial}\Lambda^{\mu_{1}...\mu_{r-1}},\label{ML-FTripletGaugeT}\\
\delta\lambda^{\mu_{1}...\mu_{r-2}} & =\partial_\nu\Lambda^{\nu\mu_{1}...\mu_{r-2}}.\nonumber
\end{align}
 The field equations \eqref{ML-FTripletEqs} follow from the flat-space action which we present in the form similar to that given in \cite{Francia:2002pt}
 \begin{align}\label{flatmetricFaction}
 S=&\int d^{D}x\left[i\bar{\chi}^{\,\mu(r-1)}\cancel{\partial}\chi_{\,\mu(r-1)}+\bar{\chi}_{\,\mu(r-1)}\partial_\nu\Psi^{\nu \,\mu(r-1)}
 +\partial_\nu\bar{\Psi}^{\nu \,\mu(r-1)}\chi_{\,\mu(r-1)}+\frac{i}{r}\bar{\Psi}^{\,\mu(r)}\cancel{\partial}\Psi_{\,\mu(r)}\right.\nonumber\\&\left.
 +(r-1)\left(-i\bar{\lambda}^{\,\mu(r-2)}\cancel{\partial}\lambda_{\,\mu(r-2)}+\partial_\nu\bar{\chi}^{\nu \,\mu(r-2)}\lambda_{\,\mu(r-2)}+\bar{\lambda}_{\,\mu(r-2)}\partial_\nu\chi^{\nu \,\mu(r-2)}\right)\right]\,,
 \end{align}
 where, for brevity, the collective index $\mu(k)$ stands for $k$ symmetrized indices $\mu_1\ldots \mu_k$.

 The construction of the metric-like Lagrangian formulation of fermionic higher-spin triplets in AdS spaces, however, encountered difficulties \cite{Sagnotti:2003qa,Buchbinder:2007ak} and has not been accomplished by now. In what follows we will show how this puzzle is resolved by deriving the metric-like action for the fermionic higher-spin triplets in AdS from their frame-like Lagrangian formulation.

\subsection{Frame-like formulation\label{sec:FLFTriplets}}

In the frame-like formulation \cite{Sorokin:2008tf} the fermionic triplet is associated with a 1-form that takes values in the space of symmetric tensor-spinors:
\[
\psi_\alpha^{a_{1}...a_{r-1}}=dx^\mu\psi_{\alpha \,\mu}^{a_{1}...a_{r-1}}\,.
\]
This form is not subject to any gamma-trace conditions\footnote{
Remember that the frame-like fermionic field transforming under an irreducible representation of the Lorentz group of a  spin $s=r+\frac{1}{2}$
is gamma-traceless \cite{Aragone:1980rk,Vasiliev:1986td,Vasiliev:1987tk}
\[
\gamma_{b}\psi_{\mu}^{a_{1}...a_{r-2}b}=0\implies\eta_{bc}\psi_{\mu}^{a_{1}...a_{r-3}bc}=0.
\]
}.

In flat space the gauge transformation of $\psi^{a_{1}...a_{r-1}}$ has the following form
\begin{equation}\label{fsgt}
\delta\psi^{a_{1}...a_{r-1}}=d\xi^{a_{1}...a_{r-1}}-dx_{b}\xi^{a_{1}...a_{r-1},b},
\end{equation}
where the tensor-spinor gauge parameters $\xi^{a_{1}...a_{r-1}}$ and
$\xi^{a_{1}...a_{r-1},b}$ are zero forms that remove from $\psi_{\mu}^{a_{1}...a_{r-1}}$
the unphysical degrees of freedom. The second parameter is associated with a connection-like
1-form $\psi_{\mu}^{a_{1}...a_{r-1},b}$.
\if{}
In general, as in the bosonic
case, we have a tower of $r-1$ of such forms $\psi^{a_{1}...a_{r-1},b_{1}...b_{p}}$
($p\leq r-1$) and their respective gauge parameters $\xi^{a_{1}...a_{r-1},b_{1}...b_{p}}$.
\fi
The connection field is auxiliary and expressed (modulo the pure gauge degrees of freedom) as a function
of $\psi_{\mu}^{a_{1}...a_{r-1}}$ via the torsion-like constraint
\begin{equation}
{\mathcal T}^{a_{1}...a_{r-1}}= d\psi^{a_{1}...a_{r-1}}-dx_{c}\wedge\psi^{a_{1}...a_{r-1},b}=0,\label{flatFTripletTorsionsConstraints}
\end{equation}
The torsion is invariant under the gauge transformations \eqref{fsgt} and that of $\psi^{a_{1}...a_{r-1},b}$
\begin{equation}\label{flatFTripletGaugeT}
\delta\psi^{a_{1}...a_{r-1},b}=d\xi^{a_{1}...a_{r-1},b}-dx_{c}\xi^{a_{1}...a_{r-1},bc}.
\end{equation}
For the consistency of the construction (see Section \ref{sec:FLFTripletAction}), while the higher-spin vielbein is unconstrained, the connection $\psi^{a_{1}...a_{r-1},b}$ and the gauge parameters
$\xi^{a_{1}...a_{r-1},b}$ and $\xi^{a_{1}...a_{r-1},bc}$ should satisfy the following (gamma-)trace constraints
\begin{gather}
\gamma_{b}\psi^{a_{1}...a_{r-1},b}=0,\quad\eta_{bc}\psi^{a_{1}...a_{r-2}b,c}=0,\label{flatFTripletConstraints}\\
\label{flatXiConstraints}
\gamma_{b}\xi^{a_{1}...a_{r-1},b}=0,\quad\eta_{bc}\xi^{a_{1}...a_{r-2}b,c}=0, \\
\gamma_{b}\xi^{a_{1}...a_{r-1},bc}=0,\quad\eta_{db}\xi^{a_{1}...a_{r-2}d,bc}=0.\nonumber
\end{gather}

The metric-like triplet fields $\Psi,\chi$ and $\lambda$ introduced in Section
$\ref{sec:MLFTriplets}$ are related to $\psi_{\mu}^{a_{1}...a_{r-1}}$ by the following identifications motivated by the form of the gauge transformations $\eqref{ML-FTripletGaugeT}$
\begin{equation}\label{FTripletIds}
\hat\Psi^{a_{1}...a_{r}}\equiv\delta^{\mu(a_{r}}\psi_{\mu}^{a_{1}...a_{r-1})},\quad\hat\chi^{a_{1}...a_{r-1}}\equiv i \gamma^{\mu}\psi_{\mu}^{a_{1}...a_{r-1}},\quad\hat \lambda^{a_{1}...a_{r-2}}\equiv\delta_{a_{r-1}}^{\mu}\psi_{\mu}^{a_{1}...a_{r-1}},
\end{equation}
where we introduced ``hatted'' quantities which differ from $\Psi,\chi$ and $\lambda$ by a total trace, as we will explain now\footnote{The imaginary unit factor $i$ in the definition of $\hat\chi$ is introduced because in our conventions when $D=4$ the gamma-matrices are purely imaginary in the Majorana representation.}.

The splitting  \eqref{FTripletIds} manifests the fact that the representation in which $\psi^{a_{1}...a_{r-1}}$
sits is not irreducible, as it may have $\gamma_a$- and $\eta_{ab}$-traces.
As a result, this field contains physical states with spins going
down from $s=r+\frac{1}{2}$ to $3/2$, while its spin 1/2 state is a pure gauge.
This is due to the fact that the three fields \eqref{FTripletIds}, being all derived from
$\psi_{\mu}^{a_{1}...a_{r-1}}$, are not completely independent. Indeed, let us define the complete
trace $\tilde{\mathbb T}$ of a tensor-spinor ${\mathbb T}^{a_{1}...a_{r}}$ as
\begin{equation}\label{trace}
\tilde{\mathbb T}=\begin{cases}
\eta_{a_{1}a_{2}}\cdot\cdot\cdot\eta_{a_{r-1}a_{r}}{\mathbb T}^{a_{1}...a_{r}} & \mbox{ if \ensuremath{r}\,is even}\\
\eta_{a_{1}a_{2}}\cdot\cdot\cdot\eta_{a_{r-2}a_{r-1}}i\gamma_{a_{r}}{\mathbb T}^{a_{1}...a_{r}} & \mbox{if \ensuremath{r}\,is odd}.
\end{cases}
\end{equation}
Then, in the metric-like description, the equations of motion imply that
for $s=r+\frac{1}{2}$
\begin{equation}\label{even}
\begin{cases}
\cancel{\partial}\left(\tilde{\Psi}-r\tilde{\lambda}\right)=0 & \mbox{ if \ensuremath{r} is even}\\
\cancel{\partial}\left(\tilde{\Psi}-\tilde{\chi}-(r-1)\tilde{\lambda}\right)=0 & \mbox{if \ensuremath{r}\,is odd}\,.
\end{cases}
\end{equation}
Their form allows us to identify the spin $\frac{1}{2}$ field as
\begin{equation}
\rho\equiv\begin{cases}
\tilde{\Psi}-r\tilde{\lambda} & \mbox{ if \ensuremath{r} is even}\\
\tilde{\Psi}-\tilde{\chi}-(r-1)\tilde{\lambda} & \mbox{ if \ensuremath{r} is odd}.
\end{cases}\label{flatSpin12Field}
\end{equation}
In the frame-like formulation, by virtue of \eqref{FTripletIds}, we find that  $\rho\equiv0$ .

This analysis tells us that the triplet fields $\eqref{FTripletIds}$ obtained from
$\psi_{\mu}^{a_{1}...a_{r-1}}$  are not
quite the same as those appearing in the equations $\eqref{ML-FTripletEqs}$:
their complete traces do not match. Of course,
we can fix this issue by simply adding a spin-$\frac{1}{2}$ massless
field $\rho$ to the frame-like action of the theory, which we will
introduce in the next section. Then the original metric-like triplet fields ${\Psi},\chi$ and $\lambda$ are related to the fields $\eqref{FTripletIds}$ as follows
\begin{align}\label{Spin12Inclusion1}
{\Psi}^{a_{1}...a_{r}} & \equiv\hat\Psi^{a_{1}...a_{r}}+\eta^{(a_{1}a_{2}}{\cdot\cdot\cdot}\eta^{a_{r-1}a_{r})}\rho\,,\nonumber \\
{\lambda}^{a_{1}...a_{r-2}} & \equiv\hat\lambda^{a_{1}...a_{r-2}}+\eta^{(a_{1}a_{2}}{\cdot\cdot\cdot}\eta^{a_{r-3}a_{r-2})}\rho\,\,\,\,\,\,\, {\rm for}~r~{\rm even}
\end{align}
and
\begin{align}\label{Spin12Inclusion2}
{\Psi}^{a_{1}...a_{r}} & \equiv\hat\Psi^{a_{1}...a_{r}}-i\eta^{(a_{1}a_{2}}{\cdot\cdot\cdot}\eta^{a_{r-2}a_{r-1}}\gamma^{a_{r})}\rho\,\nonumber \\
{\lambda}^{a_{1}...a_{r-2}} & \equiv\hat\lambda^{a_{1}...a_{r-2}}-i\eta^{(a_{1}a_{2}}{\cdot\cdot\cdot}\eta^{a_{r-4}a_{r-3}}\gamma^{a_{r-2})}\rho\,,\\
{\chi}^{a_{1}...a_{r-1}} & \equiv\hat\chi^{a_{1}...a_{r-1}}+2\eta^{(a_{1}a_{2}}{\cdot\cdot\cdot}\eta^{a_{r-2}a_{r-1})}\rho\,\,\,\,\,\,{\rm for}~r~{\rm odd}.\nonumber
\end{align}

\subsection{Frame-like action for fermionic triplets in flat space\label{sec:FLFTripletAction}}

The frame-like action that reproduces, upon making the identifications \eqref{FTripletIds}, \eqref{Spin12Inclusion1} and \eqref{Spin12Inclusion2}, the equations $\eqref{ML-FTripletEqs}$
can be found by an ansatz motivated by some simple requirements. It should be gauge-invariant under \eqref{flatFTripletGaugeT} and
have a schematic form of a free fermion action
$i\bar{\psi}\gamma\partial\psi$, where $\gamma$ stays for a product of gamma-matrices and $\bar\psi$ is the Dirac conjugate of $\psi$. Therefore, to construct the action we use
 the gauge invariant higher-spin torsion 2-form
\begin{equation}\label{Torsion}
{\cal T}^{a_{1}...a_{r-1}}\equiv d\psi^{a_{1}...a_{r-1}}-dx_{b}\wedge\psi^{a_{1}...a_{r-1},b},
\end{equation}
which is considered to be non-zero off the mass shell (compare with \eqref{flatFTripletTorsionsConstraints}).
Then, terms like $i\bar{\psi}\gamma\wedge{\cal T}$
are gauge invariant under the transformations of $\psi$ but not $\bar{\psi}$.
The most general Lorentz-invariant action constructed
using such 3-forms is \cite{Sorokin:2008tf}
\begin{align}
S_{{\cal T}} & =i\int dx^{a_{1}}\wedge\cdot\cdot\cdot\wedge dx^{a_{D-3}}\wedge{\varepsilon}_{a_{1}...a_{D-3}cdf}\left(\bar{\psi}^{b_{1}...b_{r-1}}\wedge\gamma^{cdf}{\cal T}_{b_{1}...b_{r-1}}\right.\label{flatFTripletActionWithTorsion}\\
 & \qquad\qquad\qquad\qquad\qquad\qquad\qquad\qquad\qquad\qquad\left.+c\bar{\psi}^{b_{1}...b_{r-2}c}\wedge\gamma^{d}{\cal T}_{\;b_{1}...b_{r-2}}^{f}\right),\nonumber
\end{align}
where $\varepsilon_{a_{1}...a_{D}}$ is the $D$-dimensional completely
antisymmetric Levi-Civita tensor and $c$ is an arbitrary constat.

This action  has three issues to tackle. It is not invariant under
the gauge transformations of $\bar{\psi}$, it
is not real and contains the auxiliary field $\psi^{a_{1}...a_{r-1},b}$,
that should be completely determined by $\psi^{a_{1}...a_{r-1}}$
through $\eqref{flatFTripletTorsionsConstraints}$. These three issues are
fixed by choosing a proper $c$. Indeed, it is possible
to show that for $c=-6(r-1)$ and if $\psi^{a_{1}...a_{r-1},b}$ is constrained as in \eqref{flatFTripletConstraints},
all the terms proportional to this auxiliary field disappear. Then
$\eqref{flatFTripletActionWithTorsion}$ becomes simply
\begin{align}\label{flatFTripletAction}
S & =i\int dx^{a_{1}}\wedge\cdot\cdot\cdot\wedge dx^{a_{D-3}}\wedge{\varepsilon}_{a_{1}...a_{D-3}cdf}\left(\bar{\psi}^{b_{1}...b_{r-1}}\wedge\gamma^{cdf}d\psi_{b_{1}...b_{r-1}}\right.\\
 & \qquad\qquad\qquad\qquad\qquad\qquad\qquad\qquad\left.-6(r-1)\bar{\psi}^{b_{1}...b_{r-2}c}\wedge\gamma^{d}d\psi_{\;b_{1}...b_{r-2}}^{f}\right).\nonumber
\end{align}
Integrating $\eqref{flatFTripletAction}$ by parts we can turn it (modulo total derivatives) into
its complex conjugate, so  $\eqref{flatFTripletAction}$
satisfies the reality condition. Moreover, due to our choice of $c$,
in the Hermitian conjugate version of $\eqref{flatFTripletAction}$
we can restore the (vanishing) terms proportional to $\bar{\psi}^{a_{1}...a_{r-1},b}$
and rewrite the action as
\begin{align*}
S_{\bar{{\cal T}}} & =i\int dx^{a_{1}}\wedge\cdot\cdot\cdot\wedge dx^{a_{D-3}}\wedge{\varepsilon}_{a_{1}...a_{D-3}cdf}\left(\bar{{\cal T}}^{b_{1}...b_{r-1}}\wedge\gamma^{cdf}\psi_{b_{1}...b_{r-1}}\right.\\
 & \qquad\qquad\qquad\qquad\qquad\qquad\qquad\qquad\left.-6(r-1)\bar{{\cal T}}^{b_{1}...b_{r-2}c}\wedge\gamma^{d}\psi_{\;b_{1}...b_{r-2}}^{f}\right),
\end{align*}
which is therefore equivalent to $\eqref{flatFTripletActionWithTorsion}$
up to total derivatives. The variation of $\eqref{flatFTripletAction}$
under gauge transformations can be then schematically rewritten as
\[
\delta S=\frac{\delta S_{{\cal T}}}{\delta\psi}\delta\psi+\delta\bar{\psi}\frac{\delta S_{\bar{{\cal T}}}}{\delta\bar{\psi}}=0.
\]
It vanishes because of the manifest gauge invariance of ${\cal T}^{a_{1}...a_{r-1}}$
and $\bar{{\cal T}}^{a_{1}...a_{r-1}}$ and provided that the gauge parameters $\xi^{a_{1}...a_{r-1},b}$ satisfy the constraints \eqref{flatXiConstraints}.

One can show \cite{Sorokin:2008tf}  that the equations of motion derived from the frame-like
action $\eqref{flatFTripletAction}$ are equivalent to the metric-like ones $\eqref{ML-FTripletEqs}$
modulo the subtlety
with the spin-$\frac{1}{2}$ field $\rho$, which can be included into the frame-like formulation by simply adding to the action the massless Dirac Lagrangian $L_{\frac 12}=i\bar\rho\gamma^m\partial_m\rho$.

\section{Fermionic higher-spin triplets in anti-de-Sitter spaces}
\setcounter{equation}0

\subsection{The frame-like formalism in AdS}

The frame-like description of the field $\psi^{a_{1}...a_{r-1}}$
can be generalized to the anti-de-Sitter space by employing a proper
$AdS$ covariant derivative in place of the flat-space derivative and by using a
local basis for the $AdS$ tangent space given by the vielbein $e^a=dx^m\,e^a_m(x)$. Actually,
to construct the covariant derivatives we will employ two kinds of connections. The one associated
with the invariance under $SO(1,D-1)$ local Lorentz transformations and containing the spin--connection 1--form $\omega^{ab}=-\omega^{ba}$ will be denoted
by
\begin{equation}
\nabla T^{a}=dT^{a}+\omega^{a}{}_{b}\wedge T^{b}\label{NablaDef}
\end{equation}
and will act on the $D$-dimensional vectorial tangent-space indices. For the spinorial
indices one can also use the connection associated to the whole symmetry
group of $AdS_{D}$, namely the isometry group $SO(2,D-1)$, which
includes, in addition to the Lorentz transformations generated by $J^{ab}$,
the non-commuting translations generated by $P^{a}$:
\[
\left[P^{a},P^{b}\right]=\Lambda J^{ab},
\]
where $\Lambda$ is a negative cosmological constant which defines the $AdS$ curvature
\begin{equation}\label{AdScurvature}
R^{ab}=d\omega^{ab}+\omega^{ac}\wedge\omega_c{}^{b}=-\Lambda e^a\wedge e^b\,.
\end{equation}
We denote this covariant derivative by
\begin{equation}
{\cal D}\psi_{\alpha}=d\psi_{\alpha}+\frac{1}{2}\omega^{ab}\,\left(J{_{ab}}\right)_{\alpha}^{\;\;\beta}\wedge\psi_{\beta}+e^{b}\,\left(P_{b}\right)_\alpha^{\;\;\beta}\wedge\psi_{\beta}\,,\label{CalDDef}
\end{equation}
where in the spinorial representation
\[
J^{ab}=\frac{1}{4}\left[\gamma^{a},\gamma^{b}\right],\quad P^{a}=-\frac{i}{2}\sqrt{-\Lambda}\gamma^{a}.
\]
Note that, in view of \eqref{AdScurvature}, when acting on the spinors the product of the two external differentials $\mathcal D$ vanishes \eqref{CalDDef}
\begin{equation}\label{DD=0}
{\mathcal D}^2\psi=0\,.
\end{equation}

When dealing with tensor-spinors we will assume that
${\cal D}$  acts as a  covariant differential on  vector indices and as $\eqref{CalDDef}$
on spinorial ones. In particular, the matrices $\gamma^a$ are annihilated by $\nabla$ but not by $\mathcal D$
\begin{equation}\label{CalDGamma}
{\cal D}\gamma^{a}=-\frac{i}{2}\sqrt{-\Lambda}e_{b}\left[\gamma^{b},\gamma^{a}\right]=-i\sqrt{-\Lambda}e_{b}\gamma^{ba}.
\end{equation}
In view of \eqref{AdScurvature} and \eqref{DD=0}, the following identity holds for the symmetric tensor-spinors
\begin{equation}\label{CalDSquared}
{\cal D}^{2}\psi^{a_1\ldots a_r}=\nabla^{2}\psi^{a_1\ldots a_r}=-\Lambda e^{(a_1}\wedge e_{b}\wedge\psi^{a_2\ldots a_r)b}\,.
\end{equation}

The AdS counterpart of the fermionic higher-spin torsion \eqref{Torsion} is defined as follows
\begin{equation}
{\cal T}^{a_{1}...a_{r-1}}\equiv{\cal D}\psi^{a_{1}...a_{r-1}}-e_{b}\wedge\psi^{a_{1}...a_{r-1},b}.\label{AdSTorsion}
\end{equation}
In virtue of \eqref{CalDSquared}, it is gauge invariant under the following AdS-deformation of
 the gauge transformations $\eqref{flatFTripletGaugeT}$
\begin{eqnarray}\label{AdSFTripletGaugeT}
\delta\psi^{a_{1}...a_{r-1}} & =&{\cal D}\xi^{a_{1}...a_{r-1}}-e_{b}\xi^{a_{1}...a_{r-1},b}\,,\\
\delta\psi^{a_{1}...a_{r-1},b} & =&{\cal D}\xi^{a_{1}...a_{r-1},b}-e_{c}\xi^{a_{1}...a_{r-1},bc}\nonumber\\
&& +\Lambda\left(e^{(a_{r-1}}\xi^{a_{1}...a_{r-2})b}-(r-1)e^{b}\xi^{a_{1}...a_{r-1}}\right),\nonumber
\end{eqnarray}
which coincide with $\eqref{flatFTripletGaugeT}$ in the flat limit
$\Lambda\rightarrow0$.

In flat space, for consistency, we required the gauge parameters to satisfy the constraints \eqref{flatXiConstraints}, which guarantee that the $\gamma$-trace of
the higher-spin vielbein of rank $r=s-\frac 12$  transforms under the gauge transformations in the same way
as a higher-spin vielbein of rank $r=s-\frac 32$. In the AdS space, in view of $\eqref{CalDGamma}$, the same requirement
leads to the following constraints on the gauge parameters
\[
\delta\left(\gamma_{b}\psi^{a_{1}...a_{r-2}b}\right)={\cal D}\Xi^{a_{1}...a_{r-2}}-e_{c}\Xi^{a_{1}...a_{r-2},c}\,.
\]
where
\[
\Xi^{a_{1}...a_{r-2}}\equiv\gamma_{b}\xi^{a_{1}...a_{r-2}b},\quad\Xi^{a_{1}...a_{r-2},c}\equiv-i\sqrt{-\Lambda}\gamma^{c}{}_{b}\xi^{a_{1}...a_{r-2}b}+\gamma_{b}\xi^{a_{1}...a_{r-2}b,c}.
\]
The symmetrization properties of the parameter $\Xi^{a_{1}...a_{r-2},c}$ impose the following
constraint on $\xi^{a_{1}...a_{r-1},b}$ which is the AdS generalization of \eqref{flatXiConstraints}
\[
\gamma_{b}\xi^{a_{1}...a_{r-1},b}=-i\sqrt{-\Lambda}\gamma^{(a_{1}}{}_{b}\,\xi^{a_{2}...a_{r-1})b}.
\]
Correspondingly, one imposes the analogous constraint on the auxiliary higher-spin connection
\begin{equation}
\gamma_{b}\psi^{a_{1}...a_{r-1},b}=-i\sqrt{-\Lambda}\gamma^{(a_1}{}_{b}\,\psi^{a_{2}...a_{r-2})b}.\label{AdSFTripletConstraint}
\end{equation}
The relations involving contractions with the metric $\eta_{ab}$ in $\eqref{flatFTripletConstraints}$
and $\eqref{flatXiConstraints}$ do not change.

\subsection{The frame-like action in AdS}

We construct the frame-like action for the reducible higher-spin fermionic field ${\psi}^{a_{1}...a_{r-1}}$ in the anti-de Sitter space in the same way as in flat space, i.e. with the use of the higher-spin torsion
$\eqref{AdSTorsion}$ and fix the coefficient by requiring that the action reduces to $\eqref{flatFTripletActionWithTorsion}$
in the  $\Lambda\rightarrow0$ limit. We thus get
\begin{align}
S_{AdS} & =i\int e{}^{a_{1}}\wedge\cdot\cdot\cdot\wedge e{}^{a_{D-3}}\wedge{\varepsilon}_{a_{1}...a_{D-3}cdf}\left(\bar{\psi}^{b_{1}...b_{r-1}}\wedge\gamma^{cdf}{\cal T}_{b_{1}...b_{r-1}}\right.\label{AdSFTripletActionWithTorsion}\\
 & \qquad\qquad\qquad\qquad\qquad\qquad\qquad\qquad\qquad\left.-6(r-1)\bar{\psi}^{b_{1}...b_{r-2}c}\wedge\gamma^{d}{\cal T}_{\;b_{1}...b_{r-2}}^{f}\right).\nonumber
\end{align}
This time the terms in this action containing the independent components of the auxiliary field ${\psi}^{a_{1}...a_{r-1},b}$ cancel each other due to the deformed constraint $\eqref{AdSFTripletConstraint}$. So, in comparison with \eqref{flatFTripletAction} the action
$\eqref{AdSFTripletActionWithTorsion}$ contains more terms than in the flat space, namely
\begin{align}\label{AdSFTripletAction}
S_{AdS} & =i\int{\varepsilon}_{a_{1},...,a_{D-3}cdf}e^{a_{1}}\cdot\cdot\cdot e^{a_{D-3}}\,\left\{ \bar{\psi}^{b_{1}...b_{r-1}}\gamma^{cdf}{\cal D}\psi_{b_{1}...b_{r-1}}\right.\nonumber \\
 & -\!6(r-1)\bar{\psi}^{b_{1}...b_{r-2}c}\gamma^{d}{\cal D}\psi_{b_{1}...b_{r-2}}^{f}\!+\!i\frac{\sqrt{-\Lambda}}{D-2}e^{c}\left[6(r-1)\bar{\psi}^{b_{1}...b_{r-2}d}\gamma^{f}\gamma^{g}\psi_{b_{1}...b_{r-2}g}\right.\\
 & +3(r-1)\left(\bar{\psi}^{b_{1}...b_{r-1}}\,\gamma^{df}\psi_{b_{1}...b_{r-1}}-\bar{\psi}^{b_{1}...b_{r-2}e}\gamma_{e}\,\gamma^{df}\gamma^{g}\psi_{b_{1}...b_{r-2}g}\right)\nonumber \\
 & \left.\left.-6(r-1)\left((r-2)\bar{\psi}^{b_{1}...b_{r-3}ed}\gamma_{e}\,\gamma^{g}\psi_{gb_{1}...b_{r-3}}^{k}-(r-1)\bar{\psi}^{b_{1}...b_{r-2}d}\,\psi_{b_{1}...b_{r-2}}^{f}\right)\right]\right\},\nonumber
\end{align}
where to simplify the appearance of the above expression we have skipped the wedge products of the differential forms.
In the form \eqref{AdSFTripletAction} the AdS action was constructed in \cite{Sorokin:2008tf} by a ``brute force'', i.e.  without the help of the gauge-invariant higher-spin torsion \eqref{AdSTorsion}.

The total gauge invariance and reality of $\eqref{AdSFTripletActionWithTorsion}$
(and therefore of $\eqref{AdSFTripletAction}$) is proven in the same
way as in Section $\ref{sec:FLFTripletAction}$ by showing that the action is equivalent (modulo total derivatives) to
the action constructed with the use of the Dirac conjugate torsion $\bar{{\cal T}}^{a_{1}...a_{r-1}}$.

As in the flat case, to include in the consideration the spin-$\frac 12$ field $\rho$ we add to the action \eqref{AdSFTripletAction} the Dirac action
\begin{equation}\label{adsdirac}
S_{\frac 12}=\int d^Dx\bar\rho \left(i \gamma^a\nabla_a+\frac{D-4}{2}\sqrt{-\Lambda}\right)\rho\,.
\end{equation}

\subsection{The metric-like action and equations of motion for fermionic triplets in AdS}
Having at hand the frame-like action \eqref{AdSFTripletAction}+\eqref{adsdirac} we are now ready to derive its metric-like counterpart by replacing in the former the higher-spin vielbein $\psi_\mu^{a_1\ldots a_{r-1}}$ with the fermionic higher-spin triplet fields defined in $\eqref{FTripletIds}$, \eqref{Spin12Inclusion1} and \eqref{Spin12Inclusion2}. This is achieved by first passing from the differential form expression for the action \eqref{AdSFTripletAction} to its component form in terms of $\psi_\mu^{a_1\ldots a_{r-1}}$ and then regrouping various terms with $\gamma^a$ and $\eta_{ab}$ contractions.
Somewhat tedious but direct calculations result in the following metric-like action
\begin{align}\label{FTripletMLAction}
S_{ML}^{AdS} & =\int d^{D}x\sqrt{-g}\left[i\bar{\chi}^{b_{1}...b_{r-1}}\cancel{\nabla}\chi_{b_{1}...b_{r-1}}+\bar{\chi}_{b_{1}...b_{r-1}}\nabla_a\Psi^{ab_{1}...b_{r-1}}\right.\nonumber \\
 &\left. +\nabla_a\bar{\Psi}^{ab_{1}...b_{r-1}}\chi_{b_{1}...b_{r-1}}+\!\frac{i}{r}\bar{\Psi}^{b_{1}...b_{r}}\cancel{\nabla}\!\Psi_{b_{1}...b_{r}}\!\right.\nonumber \\
 & \left.-\!(r-1)\!\left(i\bar{\lambda}^{b_{1}...b_{r-2}}\cancel{\nabla}\!\lambda_{b_{1}...b_{r-2}}\!-\!\nabla_a\bar{\chi}^{ab_{1}...b_{r-2}}\lambda_{b_{1}...b_{r-2}}\!-\!\bar{\lambda}_{b_{1}...b_{r-2}}\nabla_a\chi^{ab_{1}...b_{r-2}}\right)\right.\nonumber \\
 & \left.-\sqrt{-\Lambda}\left(\frac{D+2r}{2}\bar{\chi}^{b_{1}...b_{r-1}}\chi_{b_{1}...b_{r-1}}-\frac{D+2r-4}{2r}\bar{\Psi}^{b_{1}...b_{r}}\Psi_{b_{1}...b_{r}}\right.\right.\\
 & \left.\left. +(r-1)\frac{D+2r-8}{2}\bar{\lambda}^{b_{1}...b_{r-2}}\lambda_{b_{1}...b_{r-2}}+\frac{3}{2}i(r-1)\bar{\cancel{\chi}}^{b_{1}...b_{r-2}}\lambda_{b_{1}...b_{r-2}}\right.\right.\nonumber \\
 & - \left.\left.\frac{3}{2}i(r-1)\bar{\lambda}^{b_{1}...b_{r-2}}\cancel{\chi}_{b_{1}...b_{r-2}}+\frac{3}{2}i\bar{\cancel{\Psi}}^{b_{1}...b_{r-1}}\chi_{b_{1}...b_{r-1}}\right.\right.\nonumber \\
 & \left.\left.-\frac{3}{2}i\bar{\chi}^{b_{1}...b_{r-1}}\cancel{\Psi}_{b_{1}...b_{r-1}} +\bar{\cancel{\Psi}}^{b_{1}...b_{r-1}}\cancel{\Psi}_{b_{1}...b_{r-1}}\right.\right.\nonumber \\
 & \left.\left.-(r-1)\bar{\cancel{\chi}}^{b_{1}...b_{r-2}}\cancel{\chi}_{b_{1}...b_{r-2}}-(r-1)(r-2)\bar{\cancel{\lambda}}^{b_{1}...b_{r-3}}\cancel{\lambda}_{b_{1}...b_{r-3}}\right)\right]\nonumber
\end{align}
This action is gauge invariant under the AdS version of the gauge
transformations $\eqref{flatFTripletGaugeT}$
\begin{eqnarray}\label{gauget}
\delta\Psi^{a_{1}...a_{r}} & = & \nabla^{(a_{1}}\xi^{a_{2}...a_{r})}-\frac{i}{2}\sqrt{-\Lambda}\gamma^{(a_{1}}\xi^{a_{2}...a_{r})}\nonumber\\
\delta\chi^{a_{1}...a_{r-1}} & = & i\cancel{\nabla}\xi^{a_{1}...a_{r-1}}+\frac{D+2r-2}{2}\sqrt{-\Lambda}\xi^{a_{1}...a_{r-1}}-\sqrt{-\Lambda}\gamma^{(a_{1}}\cancel{\xi}^{a_{2}...a_{r-1})}\\
\delta\lambda^{a_{1}...a_{r-2}} & = & \nabla_{b}\xi^{a_{1}...a_{r-2}b}-\frac{i}{2}\sqrt{-\Lambda}\cancel{\xi}^{a_{1}...a_{r-2}},\nonumber
\end{eqnarray}
which follow from the first expression in $\eqref{AdSFTripletGaugeT}$. Note that these transformations differ from those considered in \cite{Buchbinder:2007ak} by terms containing the $\gamma$--trace of the symmetry parameter.

The  fermionic triplet equations of motion obtained by extremising this action have the following form
\begin{eqnarray}\label{AdSFTripletEqs}
\left(i\cancel{\nabla}+\frac{D+2r-4}{2}\sqrt{-\Lambda}\right)\Psi_{b_{1}...b_{r}}-\sqrt{-\Lambda}\gamma_{(b_{1}}\cancel{\Psi}_{b_{2}...b_{r})}\qquad\qquad\qquad\qquad\qquad\qquad{} \nonumber\\
 = \left(\nabla_{(b_{1}}+\frac{3}{2}i\sqrt{-\Lambda}\gamma_{(b_{1}}\right)\chi_{b_{2}...b_{r})}\,,\nonumber \\
\left(i\cancel{\nabla}\!+\!\frac{D+2r-8}{2}\sqrt{-\Lambda}\right)\!\lambda_{b_{1}...b_{r-2}}\!-\!\sqrt{-\Lambda}\gamma_{(b_{1}}\cancel{\lambda}_{b_{2}...b_{r-2})}\qquad\qquad\qquad\qquad\qquad\qquad{}\nonumber\\
= \!\nabla^a\chi_{ab_{1}...b_{r-2}}\!+\!\frac{3}{2}i\sqrt{-\Lambda}\cancel{\chi}_{b_{1}...b_{r-2}}\,,\nonumber \\
\left(i\cancel{\nabla}\!-\!\frac{D+2r}{2}\sqrt{-\Lambda}\right)\!\chi_{b_{1}...b_{r-1}}\!+\!\sqrt{-\Lambda}\gamma_{(b_{1}}\cancel{\chi}_{b_{2}...b_{r-1})}\qquad\qquad\qquad\qquad\qquad\qquad{}\\
=-\nabla^a\Psi_{ab_{1}...b_{r-1}}-\frac{3}{2}i\sqrt{-\Lambda}\cancel{\Psi}_{b_{1}...b_{r-1}} +\left(\nabla_{(b_{1}}+\frac{3}{2}i\sqrt{-\Lambda}\gamma_{(b_{1}}\!\!\right)\!\lambda_{b_{2}...b_{r-1})}\,.\nonumber
\end{eqnarray}

\subsection{($\frac 12$,$\frac 32$) doublet}
As the simplest example demonstrating basic properties of the above metric-like Lagrangian systems of fermionic fields in AdS, let us consider a doublet of fields $\chi$ and $\Psi_a$ propagating a spin $\frac 12$ and $\frac 32$. For this system the action reduces to
\begin{eqnarray}\label{32act}
S_{doublet}& =\int d^{D}x\sqrt{-g}\left[i\bar{\chi}\cancel{\nabla}\chi+\bar{\chi}\nabla_a\Psi^a +\nabla_a\bar{\Psi}^{a}\chi+{i}\bar{\Psi}^{a}\cancel{\nabla}\!\Psi_{a}\right.\nonumber \\
 & \left.-\sqrt{-\Lambda}\left(\frac{D+2}{2}\bar{\chi}\chi
 -\frac{D-2}{2}\bar{\Psi}^{a}\Psi_{a}+\frac{3}{2}i\bar{\cancel{\Psi}}\chi -\frac{3}{2}i\bar{\chi}\cancel{\Psi} +\bar{\cancel{\Psi}}\cancel{\Psi}\right)\right]
\end{eqnarray}
and the equations of motion \eqref{AdSFTripletEqs} become
\begin{eqnarray}\label{32d}
\left(i\cancel{\nabla}+\frac{D-2}{2}\sqrt{-\Lambda}\right)\Psi_{a}-\sqrt{-\Lambda}\gamma_{a}\gamma^b{\Psi}_b & =& \left(\nabla_{a}+\frac{3}{2}i\sqrt{-\Lambda}\gamma_{a}\right)\chi\,,\nonumber \\
i\nabla^a\Psi_{a}-\frac{3}{2}\sqrt{-\Lambda}\gamma^a{\Psi}_{a}&=&\left(\cancel{\nabla}\!+\!i\frac{D+2}{2}\sqrt{-\Lambda}\right)\chi.
\end{eqnarray}
These are invariant under the gauge transformations
\begin{eqnarray}\label{32g}
\delta\Psi_{a} & = & \nabla_{a}\xi-\frac{i}{2}\sqrt{-\Lambda}\gamma_{a}\xi\,,\\
\delta\chi& = & i\gamma^a{\nabla}_a\xi+\frac{D}{2}\sqrt{-\Lambda}\xi\,.\nonumber
\end{eqnarray}
By taking linear combinations of the equations \eqref{32d} we get the disentangled conventional equations of motion for the gauge-invariant dynamical spin-1/2 field $\rho=\chi-i\gamma^a\Psi_a$ and the dynamical spin-3/2 field $\psi_a=\Psi_a-\frac{i}{D-2}\gamma_a\rho$
\begin{eqnarray}\label{32d1}
&\left(i\gamma^a\nabla_{a}+\frac{D-4}{2}\sqrt{-\Lambda}\right)\rho=0\,,&\nonumber\\
&\gamma^{b}{\mathcal D}_{[b}\psi_{a]}=\gamma^b(\nabla_{[b}-\frac i2\sqrt{-\Lambda}\gamma_{[b})\psi_{a]}=0.&
\end{eqnarray}
Let us now compare the equations \eqref{32d} with equations for a ($\frac 12$,$\frac 32$) doublet proposed in \cite{Sagnotti:2003qa}. The latter have the following form
\begin{eqnarray}\label{32st}
&\left(i\cancel{\nabla}+\frac{D-2}{2}\sqrt{-\Lambda}\right)\Psi_{a}+\frac{\sqrt{-\Lambda}}{2}\gamma_{a}\gamma^b{\Psi}_b  = \nabla_{a}\chi\,,&\nonumber \\
&i\nabla^a\Psi_{a}+\frac{D-1}{2}\sqrt{-\Lambda}\gamma^a{\Psi}_{a}=\cancel{\nabla}\chi.&
\end{eqnarray}
These equations are also invariant under the gauge transformations \eqref{32g} but, as one can see, they differ from \eqref{32d}. As was shown in \cite{Sagnotti:2003qa}, the consistency of \eqref{32st}  requires that
$$
\chi=i\gamma^{a}\Psi_{a},
$$
which means that the system \eqref{32st} does not contain the physical spin-1/2 field, the issue which is solved by the properly modified equations \eqref{32d}.

\section{Conclusion and outlook}
As was shown in \cite{Sorokin:2008tf}, in the frame-like formulation, the triplet fields are endowed with a geometrical meaning of higher-spin vielbeins and connections transforming under higher-spin local symmetries. This allows one to determine in a conventional way gauge-invariant higher-spin torsion and curvatures and use them for the construction of simply-looking frame-like actions for these systems both in flat and AdS spaces.

Starting from the frame-like action for the unconstrained fermionic higher-spin vielbein and the spin-1/2 field in AdS space, and using the splitting of this vielbein into the metric-like triplet of fermionic fields, we have resolved a long-standing issue of the construction of the metric-like Lagrangian description of the fermionic triplets in AdS spaces.

Having now at our disposal the metric-like Lagrangian formulation for bosonic and fermionic triplets in AdS, one can analyze the BRST structure associated with their gauge transformations and equations of motion, and use the obtained form of the BRST charge for comparing it with that of \cite{Sagnotti:2003qa} and studying whether and how the triplets in AdS may arise from the quantization of strings in AdS in a tensionless limit. Since in AdS one may play with two parameters, the string tension and the AdS radius, the tensionless limit of AdS strings may avoid singularities of its flat space counterpart. In this respect it will be interesting to revise the procedure and results of taking a tensionless limit of a bosonic AdS string considered in \cite{Bonelli:2003kh}.

As another direction of research, one can proceed with studying interactions of higher-spin triplet fields (cubic and quartic vertices, current exchanges,  etc.) along lines put forward in \cite{Fotopoulos:2008ka,Fotopoulos:2009iw,Metsaev:2012uy}. One of the advantages here is that reducibility of triplet systems can make things simpler, since a single triplet vertex contains a number of vertices of irreducible higher-spin fields.

Finally, one may study whether the minimal Vasiliev theory \cite{Vasiliev:1990en}, describing interacting fields of the even spins from 0 to infinity (which can be viewed as a single higher-spin ``triplet'' with $s=\infty$), could be extended to an interacting theory of infinite sets of ``triplets'', as might happen in string theory. To this end one will need to generalize the unfolded machinery to deal with weak trace and gamma-trace conditions (like in \eqref{flatFTripletConstraints} and \eqref{flatXiConstraints}). First steps in this direction were made in \cite{Sorokin:2008tf}.

\section*{Acknowledgements}
The authors are grateful to Dario Francia, Ruslan Metsaev, Bo Sundborg, Mirian Tsulaia and Mikhail Vasiliev for useful discussions and comments.  The work of D.~S.~was partially supported by the Russian Science Foundation grant 14-42-00047 in association with Lebedev Physical Institute and by the Australian Research Council (ARF) Discovery Project grant DP160103633. The work of F. A. was supported by INFN and Scuola Galileiana di Studi Superiori of Padua.

\section*{Appendix. Notation and conventions}
The signature of the $D$-dimensional space-time metric is chosen to be almost minus $(+,-,\ldots,-)$.
The Greek letters
$\mu,\nu,...$ denote world indices associated with space-time coordinates $x^\mu$. The Latin letters $a,b,c...$ label the components of tangent-space tensors. The world indices are converted
into the tangent space ones by means of the vielbein $e_{\mu}^{a}(x)$, which is just the unit matrix $\delta_\mu^a$ in flat space-time.

Different groups of symmetric indices are separated by commas. Each group corresponds to a row in the Young tableau associated to the representation the tensor sits in. For example,\ytableausetup{boxsize=1.1em}
\[
\psi^{a_1a_2\ldots a_m,b_1\ldots b_n}, \qquad (n\leq m)
\]
is a tensor whose symmetry properties are defined
by the Young tableau
$$
\ytableausetup{centertableaux}
\begin{ytableau}
a_1 & a_2 &\ldots & a_m \\
b_1 & \ldots & b_n
\end{ytableau}
$$
i.e. $\psi^{(a_1a_2\ldots a_m,b_1)\ldots b_n}=0$.

Symmetrizations of indices are not weighted and are denoted with round
brackets, e.g.
\[
A^{(a_{1}}B^{a_{2}a_{3})}\equiv A^{a_{1}}B^{a_{2}a_{3}}+A^{a_{2}}B^{a_{1}a_{3}}+A^{a_{3}}B^{a_{2}a_{1}}.
\]
We also use the short-hand notation for contractions involving  $\gamma$ matrices, e.g.
\[
\cancel{\psi}^{a_{1}...a_{n}}\equiv\gamma_{a_{n+1}}\psi^{a_{1}...a_{n}a_{n+1}},\quad \cancel{\partial}\equiv\gamma_{\mu}\partial^{\mu}\,.
\]
The gamma--matrices obey the Clifford algebra
$$
\gamma^a\gamma^b+\gamma^b\gamma^a=2\eta^{ab}\,.
$$
In $D=4$ the gamma-matrices are purely imaginary in the Majorana representation.

\if{}
\bibliographystyle{utphys}
\bibliography{references}
\fi

\providecommand{\href}[2]{#2}\begingroup\raggedright\endgroup

\end{document}